\documentclass[aps,prf,longbibliography,onecolumn]{revtex4-2}

\usepackage{graphicx}
\usepackage{amsmath}
\usepackage{amssymb}
\usepackage{bm}
\usepackage{upgreek}
\usepackage{hyperref}
\hypersetup{colorlinks=true,allcolors=blue}
\usepackage{xcolor}

\usepackage[utf8]{inputenc}  % Pour les caractères accentués
\usepackage[T1]{fontenc}      % T1, pas TU !
\newcommand{\orcid}[1]{%
  \href{https://orcid.org/#1}{\raisebox{-0.05em}{\includegraphics[height=1em]{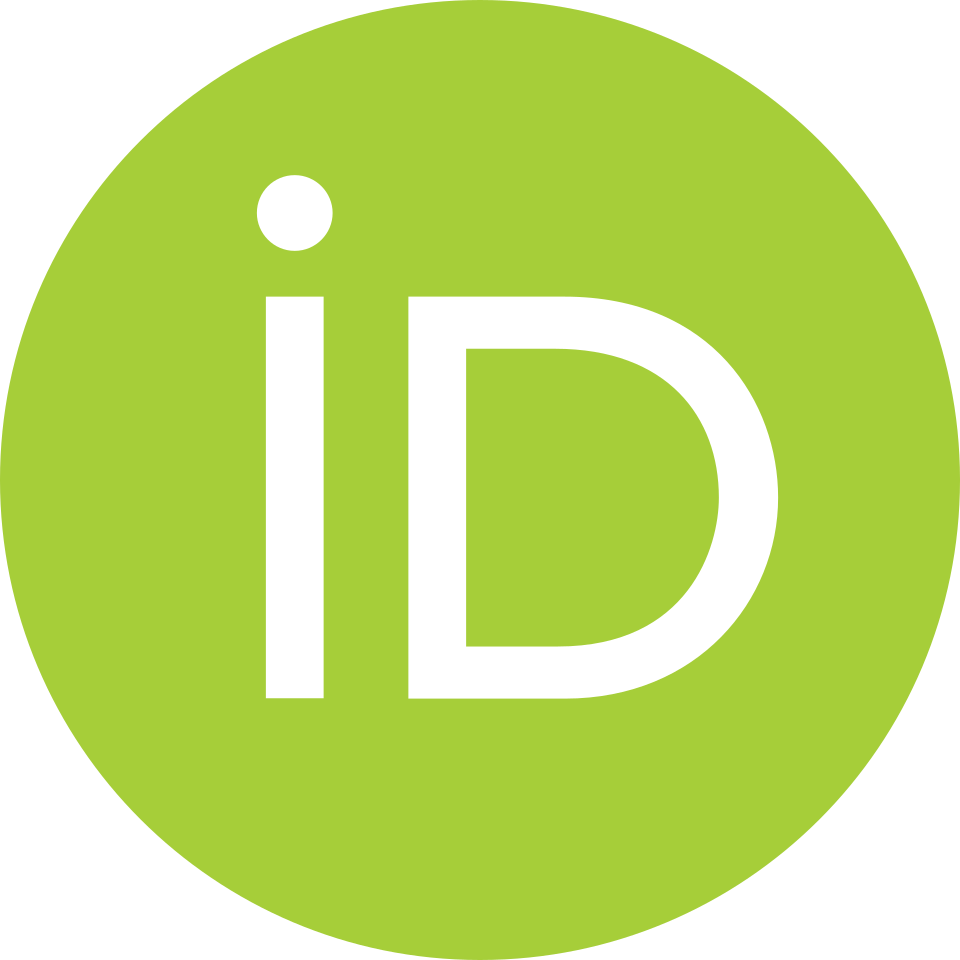}}}%
}

\begin{document}

\title{Free-surface deformations induced by three-dimensional turbulence}

\author{Micha\"el Berhanu\orcid{0000-0001-9099-2135}}
\affiliation{Universit\'e Paris Cit\'e, CNRS, MSC Laboratory, UMR 7057, F-75013 Paris, France}

\author{Eric Falcon\orcid{0000-0001-9640-9895}}
\email{eric.falcon@u-paris.fr}
\affiliation{Universit\'e Paris Cit\'e, CNRS, MSC Laboratory, UMR 7057, F-75013 Paris, France}

\date{\today}

\begin{abstract}
We report the experimental characterization of free-surface deformations generated by three-dimensional homogeneous and isotropic turbulence. Using Fourier transform profilometry in a jet-forced turbulent tank, we perform spatiotemporal measurements of the surface elevation field over a wide range of turbulence intensities. The standard deviation of surface deformations scales linearly with subsurface velocity fluctuations. The spectra of surface deformations highlight the coexistence of two mechanisms: transient coherent structures (e.g., upwelling) contributing to the low-frequency, large-scale spectral components, and a passive response to subsurface turbulent pressure fluctuations responsible for the power-law spectral scaling. The wavenumber and frequency spectra of surface deformations exhibit similar power-law exponents ($-2.5$), suggesting the advection of turbulent structures at the free surface. We develop a linear response model based on the transfer function from the free surface to turbulent pressure fluctuations, incorporating wave-turbulent damping. The model successfully predicts the main features of the turbulent surface: spatiotemporal spectrum shape, similar spectrum power-law exponents ($-7/3$), and dominance of passive response over wave generation. These findings provide new insights into free-surface turbulence in regimes where turbulent velocities remain below the surface-breaking threshold.

\end{abstract}

\maketitle

\section{Introduction}
The interaction between turbulent flows and free surfaces is fundamental to numerous geophysical and industrial processes, from ocean dynamics and atmospheric exchanges to chemical engineering applications. Understanding how subsurface turbulence deforms a free interface remains challenging due to the complex coupling between turbulent velocity and pressure fields and the surface boundary conditions imposed by gravity and surface tension. A central question is whether surface deformations result primarily from resonant generation of propagating waves or from passive response to subsurface turbulent pressure fluctuations.

Previous theoretical work has identified two main mechanisms for surface deformation generation by turbulence in air or in water. First, Phillips~\cite{Phillips1957} proposed that turbulent wind pressure fluctuations could resonate with surface waves, generating propagating gravity-capillary waves when the mean flow velocity matches the wave phase velocity. This nonlinear resonance mechanism has been refined by work showing the importance of matching both velocity and length scales~\cite{Teixeira2006}. Second, in nonresonant conditions where this matching criterion is not satisfied, turbulence deforms mainly the free-surface by the passive response to pressure fluctuations~\cite{Brocchini2001}. Recent developments have provided linear transfer function approaches to model this passive response~\cite{Perrard2019}. Numerical investigations using direct numerical simulations (DNS) have provided insights into bubble entrainment, wave breaking dynamics, and the interplay between surface tension and turbulence~\cite{Deike2022,Calado2025}. Studies of gas transfer rates using active grid turbulence have demonstrated that vertical turbulence intensity correlates strongly with interfacial exchange, while surface deformations were found to play a minor role~\cite{Bullee2024}. Experimental studies of breaking bores provide a classification of complex air-water interfacial features~\cite{Wuthrich2021}, refining the initial Brocchini and Peregrine's classification~\cite{Brocchini2001}. Free-surface behaviors of shallow turbulent flows have been reviewed~\cite{Muraro2021}, emphasizing the need for better links between bulk flow parameters and surface patterns.

When turbulence occurs only in the liquid, and the air effect is negligible, experimental advances have improved our understanding of free-surface turbulence, although results depend on the flow configuration. Studies using grid-generated turbulence in open channel flows have observed various surface patterns, including upwellings, downdrafts, and spiral eddies~\cite{Loewen1986,Kumar1998}, or propagation of gravity-capillary waves without highlighting the mechanism of their generation~\cite{Savelsberg2009}. In configuration with zero mean flow, recent experimental work using three-dimensional (3D) turbulence random-jet-stirred in close tanks has provided a detailed characterization of 3D turbulence beneath a free surface~\cite{Jamin2025,Ruth2024}, confirming some theoretical predictions~\cite{Hunt1978,Teixeira2000,Magnaudet2003}. In particular, a strong anisotropy develops beneath the surface with enhanced horizontal turbulent kinetic energy, decreased vertical velocity fluctuations, and amplified large-scale horizontal velocity spectra~\cite{Jamin2025}. The role of upwellings and downwellings has also been documented~\cite{Qi2025}, reflecting visible surface features (such as scars and dimples)~\cite{Brocchini2001,AarnesJFM2025} as well as strong surface-to-bulk correlations~\cite{BabikerArXiv2025}.  Numerical simulations of a similar configuration showed low-frequency structures (upwellings and downwellings) dominating over waves~\cite{Guo2010}. %The question of which mechanism dominates under what conditions remains partially unresolved. 
Studies have also examined the interaction between forced surface waves and subsurface turbulence~\cite{Jamin2016,Gutierrez2016,Falcon2009,SmeltzerJFM2023}, revealing complex advection and scattering phenomena. In particular, turbulent flows strongly damp surface waves through advection~\cite{Gutierrez2016}. Despite these advances, comprehensive spatiotemporal measurements of purely turbulence-generated surface deformations across a wide range of turbulence intensities remain scarce, particularly in regimes where nonresonant mechanisms may dominate. Furthermore, quantitative comparisons between experimental observations and theoretical models have been limited. %of passive surface response to turbulent pressure fluctuations 

Here, we experimentally investigate free-surface deformations induced by 3D homogeneous and isotropic hydrodynamic turbulence with zero mean flow using randomly actuated jets. We observe various surface features, including scars, upwellings, and gravity-capillary waves. Using spatiotemporal measurements of the free-surface deformations across a broad parameter space, we characterize the amplitude scaling, spectral properties, and physical nature of these deformations over a wide range of turbulence levels, reaching significant surface steepness (up to 0.1). We develop a model based on the linear transfer function of the free surface responding to turbulent pressure fluctuations, incorporating realistic turbulent damping effects. Our experimental results reveal the coexistence of passive response, creating advected structures and propagating gravity-capillary waves. The passive response model captures some features of the turbulent surface, including identical power-law exponents in both spatial and temporal spectra of surface deformations. These findings complement recent advances in understanding free-surface turbulence and provide new quantitative insights into the nonresonant regime of turbulence-interface interactions.

\section{Experimental setup}

\subsection{Turbulence generation}
\begin{figure}[t!]
\centering
\includegraphics[width=0.9\textwidth]{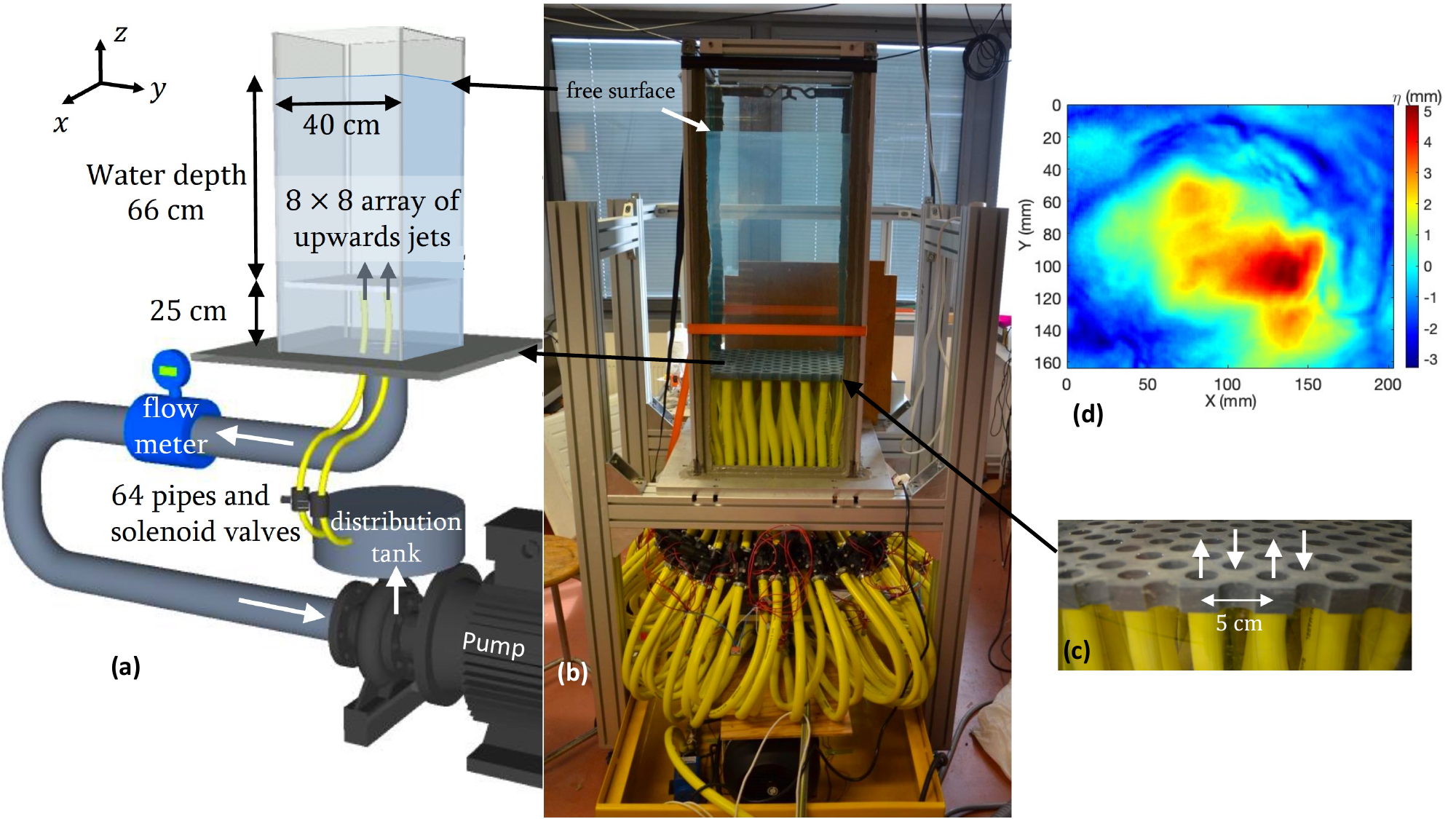}
\caption{(a) Schematic diagram of the experimental setup used to produce bulk turbulence, with almost no mean flow, within a square glass tank of side 40 cm. Water depth: 66 cm. (b) Side view of the experimental setup where 64 upward jets, arranged in a $8 \times 8$ square grid with a 5-cm spacing, are placed in the gray slab, 25 cm above the tank bottom. (c) Close-up view of the jet orifices (upward arrows) inside the slab and the suction holes (downward arrows) located at the center of each square formed by the inlet jets. (d) Typical water free-surface deformations $\eta(x,y)$ measured in a $14\times14$~cm$^2$ zone (at the tank top center), using Fourier transform profilometry (FTP), for moderate subsurface turbulence ($\sigma_{\mathcal{U}} = 5.1$~cm/s). }
\label{fig:setup}
\end{figure}

We generate three-dimensional turbulence in a square glass tank ($40\times 40$~cm$^2$, 66~cm deep) using an array of 64 vertical water jets arranged in an 8$\times$8 square grid at the tank bottom (see Fig.~\ref{fig:setup}). The jet orifices have an internal diameter 1.65~cm and are spaced 5~cm apart. A variable-flow pump supplies water through electronically controlled solenoid valves. The valves open and close randomly with controlled timing, maintaining exactly 16 active jets at any instant. This forcing creates statistically stationary, spatially homogeneous turbulence in the bulk flow with almost no mean flow~\cite{Jamin2025}. A detailed description and characterization of this experimental setup can be found in our recent publication~\cite{Jamin2025}. The flow rate is controlled via a proportional-integral-derivative (PID) feedback system, allowing precise adjustment of turbulence intensity. We characterized the turbulence properties using particle image velocimetry (PIV) measurements of the horizontal velocity field at various depths~\cite{Jamin2025}. The turbulence is homogeneous and isotropic for depths deeper than $-6$ cm. The root-mean-square velocity fluctuations $\sigma_{\mathcal{U}}$ (averaged over depth $z\in [-9,-7]$~cm and horizontal $x\in [-5, 5]$~cm) serve as our control parameter, ranging from $\sigma_{\mathcal{U}} = 0.65$ to 11~cm/s in this study. These velocity fluctuations are typically found 10 times stronger than the mean flow. The dissipation rate is $\epsilon\in [0.01,71]$~cm$^2/$s$^3$~~\cite{Jamin2025}. The integral length scale of the turbulence is approximately $L \approx 8$~cm, independent of forcing intensity~\cite{Jamin2025}. Beneath the free surface, a redistribution of turbulent kinetic energy from vertical to horizontal fluctuations is observed~\cite{Jamin2025} consistently with theoretical predictions~\cite{Hunt1978,Teixeira2000}. The turbulent Reynolds number $\mathrm{Re} \equiv \sigma_{\mathcal{U}} L/\nu$ ranges $\mathrm{Re} \in [523,8801]$ and the Taylor microscale Reynolds number $\mathrm{Re}_{\lambda} \equiv \sigma_{\mathcal{U}}^2 \sqrt{15/(\nu\epsilon)}$ ranges $[137, 558]$ with $\nu$ the kinematic viscosity of water. The influence of the flow inertia (i.e., turbulent kinetic energy) leading to surface deformations with respect to gravity, $g$, preventing deformations, is quantified by the Froude number $\mathrm{Fr} \equiv \sigma_{\mathcal{U}}/\sqrt{2gL}$ in the range $[0.52,8.8]\times 10^{-2}$. The influence of the flow inertia with respect to the surface tension is quantified by the Weber number $\mathrm{We} \equiv \sigma_{\mathcal{U}}^2L\rho/(2\gamma)$ in the range $[0.028, 8.1]$ ($\rho=1000$~kg/m$^3$ is the water density and $\gamma\simeq60$~mN/m the air-water surface tension). These parameters correspond to a domain where the free surface is not breaking and where weak turbulence, waves, and scars coexist on the surface~\cite{Brocchini2001}.

\subsection{Surface measurement technique}
We measure the free-surface elevation field $\eta(x,y,t)$ using Fourier transform profilometry (FTP)~\cite{Takeda1983,Cobelli2009,FalconARFM2022}. This non-intrusive optical technique projects a sinusoidal fringe pattern onto the free surface using a full-HD video projector (Epson EH-TW3200) positioned above the tank. Surface deformations distort the fringe pattern, which is recorded by a high-speed camera (Phantom V10) viewing the surface at an oblique angle. Fourier analysis of each image extracts the phase field, from which the surface elevation is reconstructed using a calibrated phase-to-height relationship~\cite{Cobelli2009}.

The measurement domain covers a surface $\mathcal{S}=14\times 14$~cm$^2$ at the top tank center with a spatial resolution of 0.5~mm. We acquire data at 300 frames per second for $\mathcal{T}=52$~s, providing excellent temporal resolution for analyzing both advected structures and propagating waves. To obtain good optical contrast, we seed the water surface with a controlled amount of micrometric titanium dioxide (TiO$_2$) particles~\cite{TiO2}. Previous studies have confirmed that TiO$_2$ particles do not alter the water surface tension and viscosity~\cite{Przadka2012}, ensuring that the measurements remain unaffected by the seeding particles.

\begin{figure}[t!]
\includegraphics[width=0.4\textwidth]{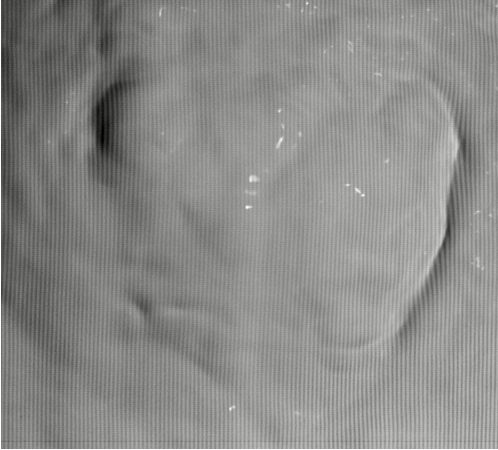}
\includegraphics[width=0.437\textwidth]{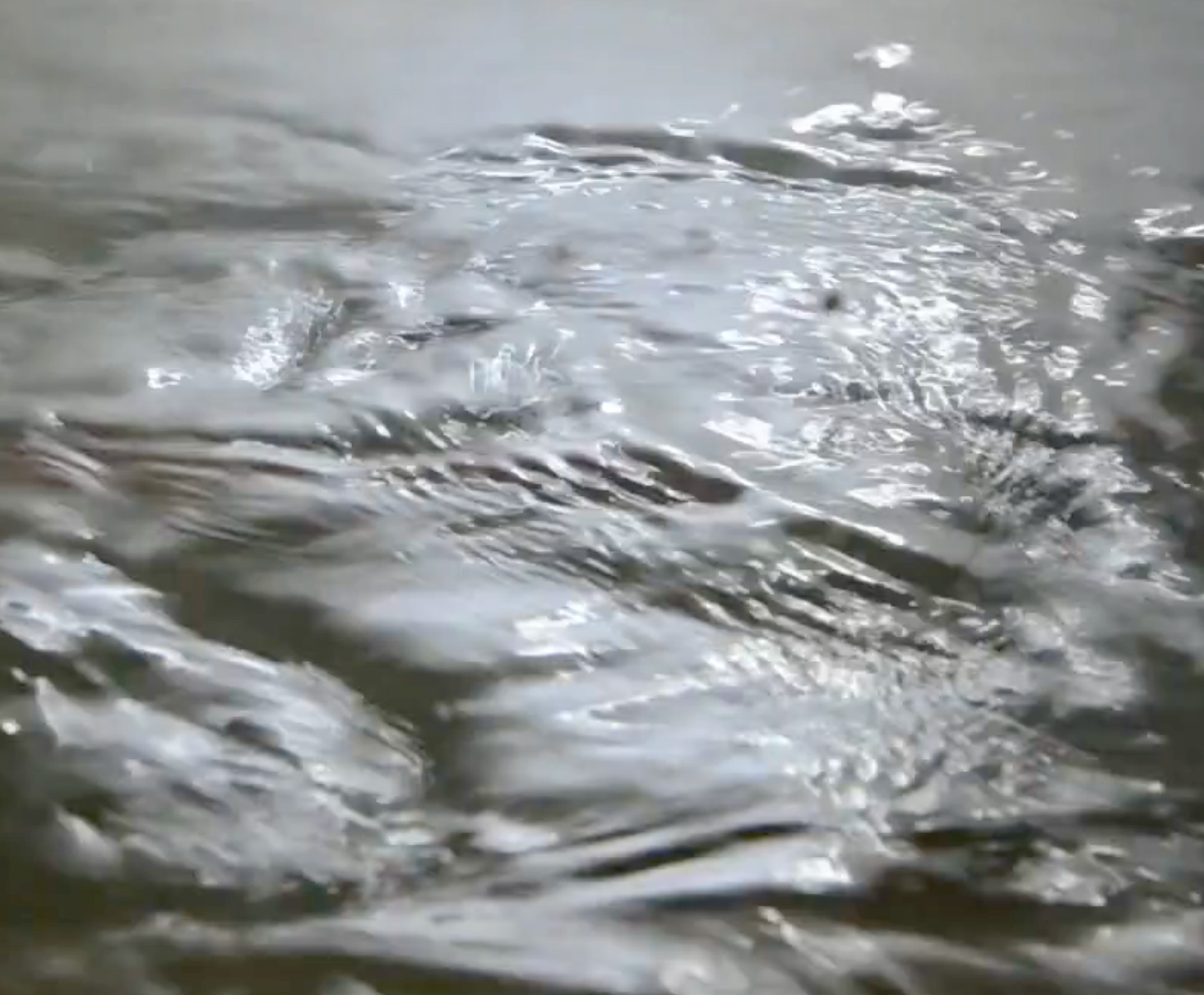}
\caption{Typical surface deformations: (Left) for gentle turbulence showing surface scarification and upwellings as imprints of subsurface turbulence (top view with fringes of the FTP method - $14\times 14$~cm$^2$), and (Right) for moderate turbulence showing scars and waves (typical size 1~cm) (oblique view).} %\rev{Which typical values of the flow rate or $\sigma_{\mathcal{U}}$?}}
\label{fig:wavesandscars}
\end{figure}

\section{Experimental results}

\subsection{Free-surface deformation amplitude}
\begin{figure}[t!]
\centering
\includegraphics[width=0.45\textwidth]{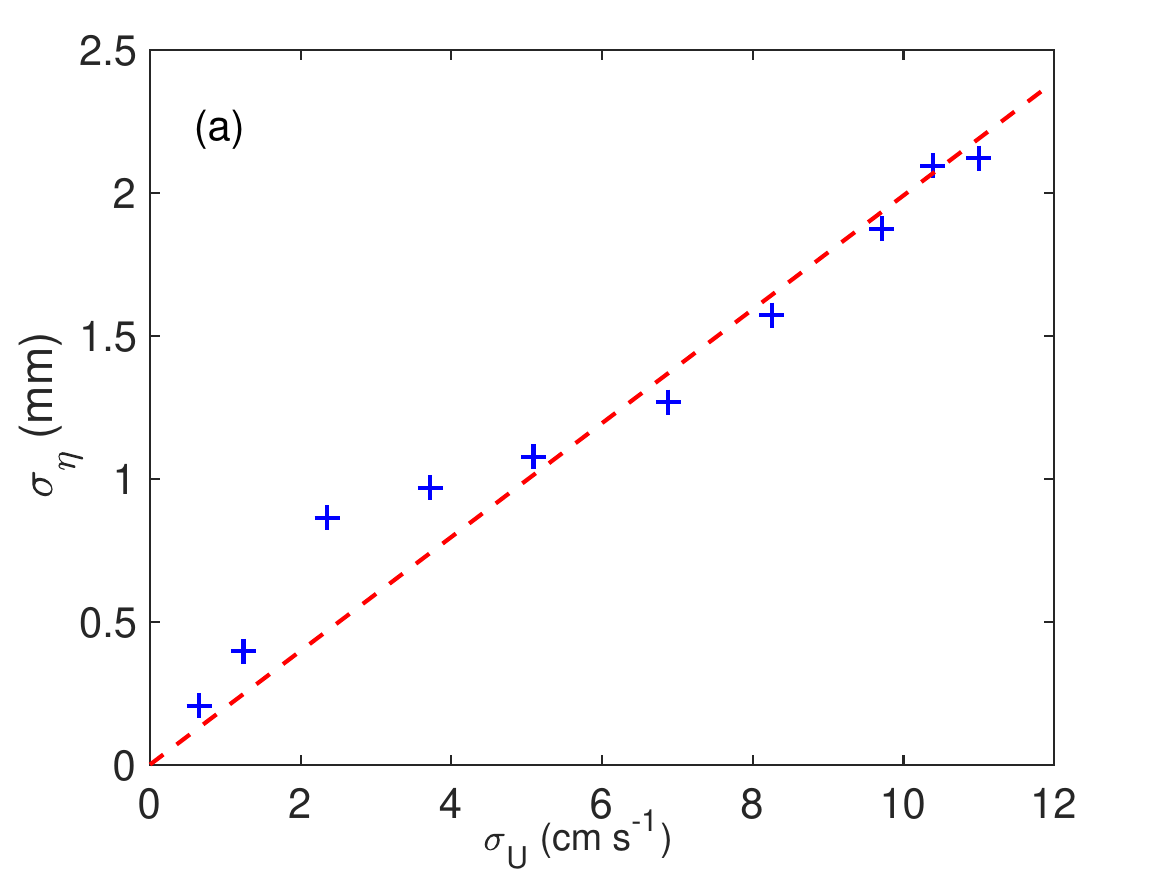}%fig_amplitude.png}
\includegraphics[width=0.47\textwidth]{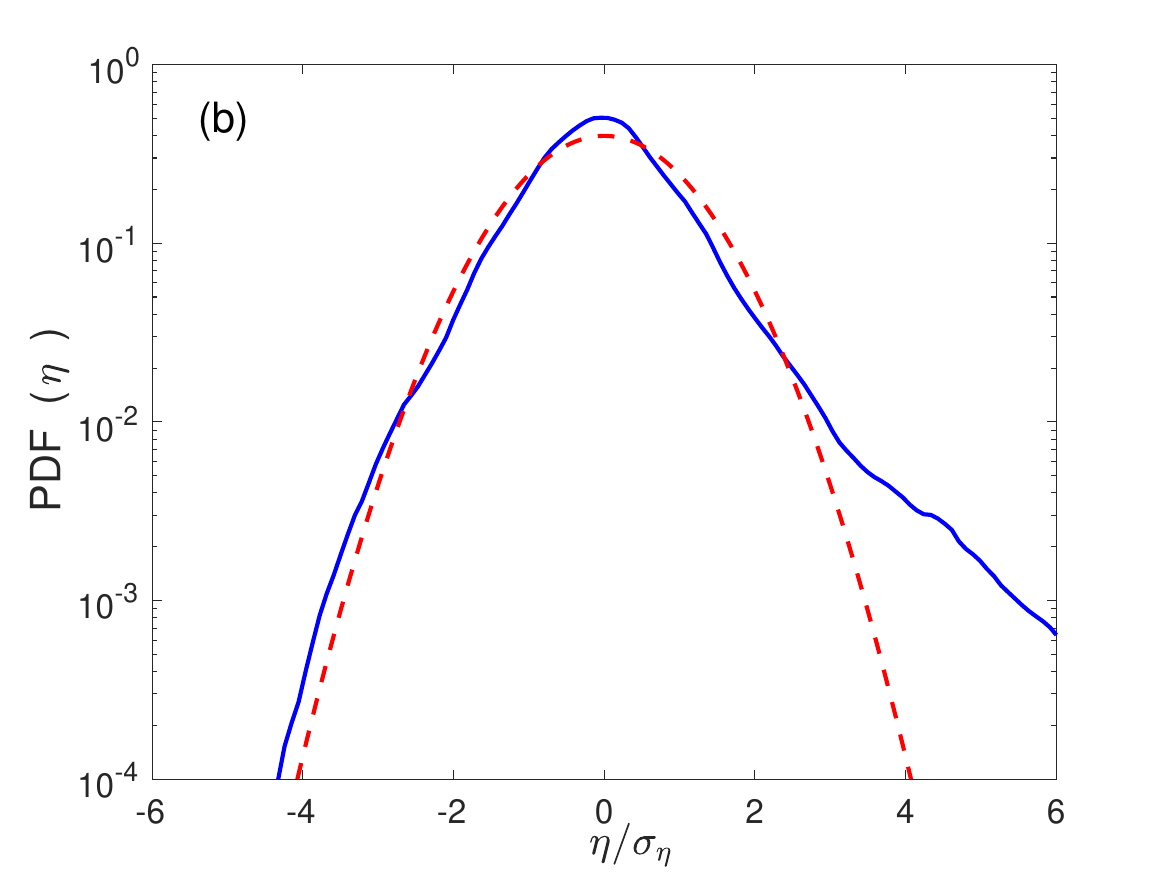}%PDF.png}
\caption{(a) Standard deviation of surface deformations $\sigma_\eta$ versus subsurface turbulent velocity fluctuations $\sigma_{\mathcal{U}}$. $\sigma_\eta$ scales linearly with turbulence intensity. The dashed red line has a slope of 0.02~s. (b) Probability density function (PDF) of the normalized surface deformations, $\eta/\sigma_{\eta}$, for $\sigma_{\mathcal{U}}=10.4$~cm/s. Dashed red line: Normal distribution.}%Mean steepness $\kappa = k_{\rm max}\sigma_\eta$ versus $\sigma_{\mathcal{U}}$. }
\label{fig:amplitude}
\end{figure}

Figure~\ref{fig:wavesandscars} shows typical surface deformations in response to subsurface turbulence. For gentle turbulence (Fig.~\ref{fig:wavesandscars}a), surface scarifications are visible as imprints of subsurface turbulence and vortices, whereas for moderate turbulence (Fig.~\ref{fig:wavesandscars}b), scars and waves are observed simultaneously (see a movie in Ref.~\cite{SuppMat}). An example of the free-surface deformation amplitude, $\eta(x,y)$, obtained by FTP is shown in Fig.~\ref{fig:setup}d and is of the order of a few mm for moderate turbulence.  The standard deviation of surface elevation $\sigma_\eta \equiv \sqrt{\langle \int_\mathcal{S}\eta(x,y,t)^2dxdy/\mathcal{S}\rangle_t}$ is displayed in Fig.~\ref{fig:amplitude}a as a function of subsurface turbulent velocity fluctuations $\sigma_{\mathcal{U}}$. The relationship is roughly linear across the entire range 
\begin{equation}   %$\sigma_\eta\equiv \sqrt{\langle \eta(t)^2\rangle_t}$ 
\sigma_\eta \approx K \sigma_{\mathcal{U}}
\label{EtaUscaling}
\end{equation}
with $K=0.02$~s and a maximum value of $\sigma_\eta \approx 2$~mm. This linear scaling suggests that the mechanism generating surface deformations remains consistent across all turbulence levels investigated. We observe that the prefactor $K$ is of the same order of magnitude as the timescale obtained from dimensional analysis $\sqrt{L/g} \approx 0.09$~s. We define a mean steepness $\kappa \equiv k_{\rm max}\sigma_\eta$, where $k_{\rm max} = 46$~m$^{-1}$ is the wavenumber of maximum energy in the spatial spectrum of deformation amplitudes (discussed below in Sect.~\ref{spacespec}). The steepness reaches $\kappa \approx 0.1$ at the highest forcing, representing significant surface deformations. The absence of any threshold behavior in the $\sigma_\eta$ vs. $\sigma_{\mathcal{U}}$ relationship indicates that surface deformation generation does not require a critical turbulence intensity. Note that when the turbulence intensity is further increased, transient humps appear on the surface beneath jets, a case which will not be studied here. Figure~\ref{fig:amplitude}b shows the probability density function (PDF) of the surface deformations $\eta(x,y,t)$, averaged over space and time,  normalized by its rms value, $\sigma_{\eta}$, for strong turbulence. The PDF is skewed with more probable positive deformations than negative ones, as a signature of upwelling events.

\begin{figure}[t!]
\centering
\includegraphics[width=0.45\textwidth]{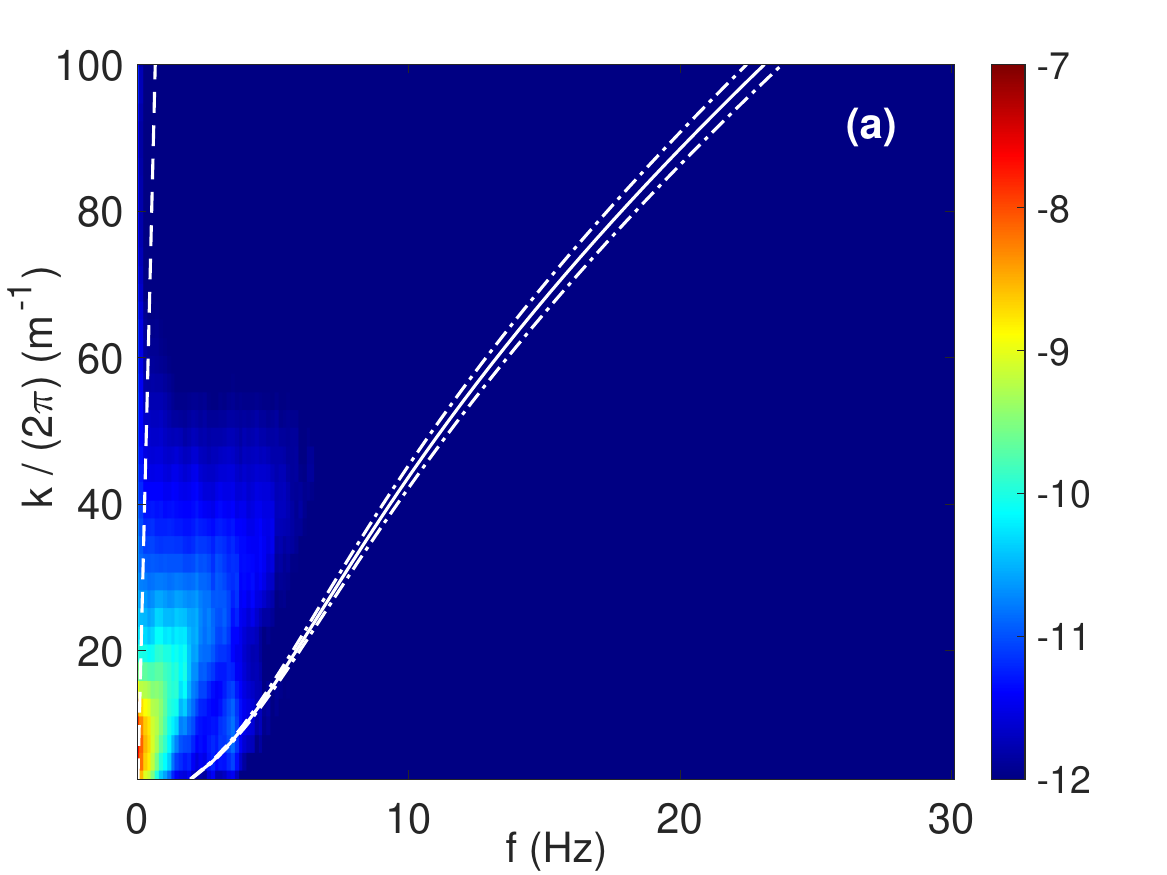} %spatiotemporal_weak.png}
\includegraphics[width=0.45\textwidth]{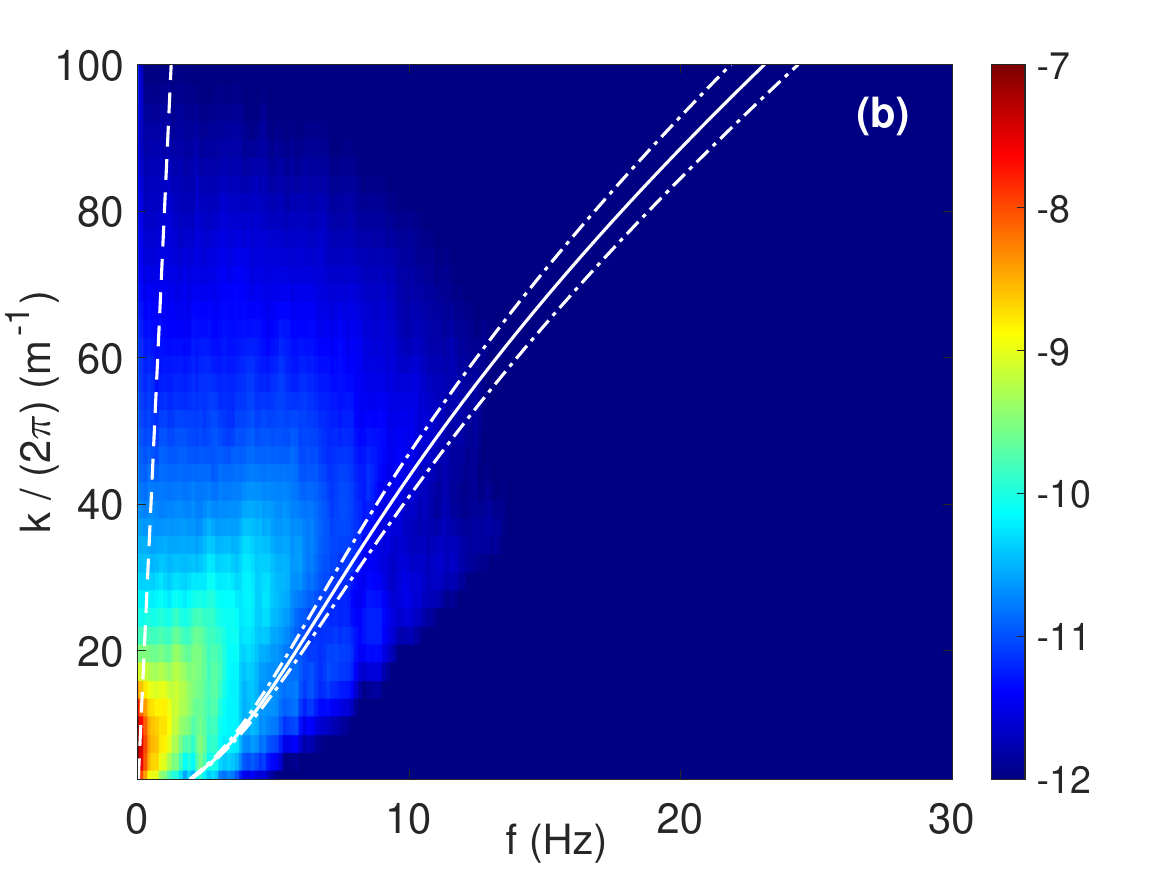}\\
\includegraphics[width=0.45\textwidth]{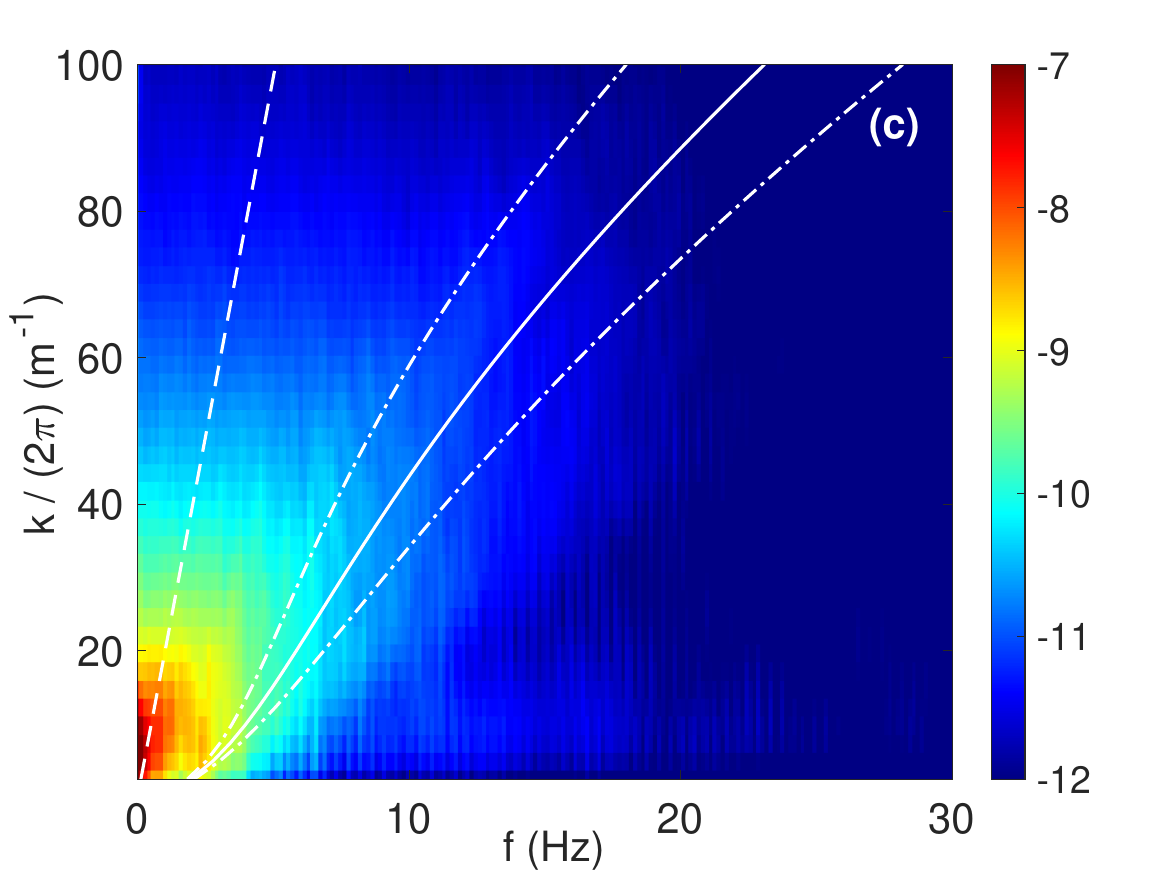}%fig5a.pdf spatiotemporal_strong.png}
\includegraphics[width=0.45\textwidth]{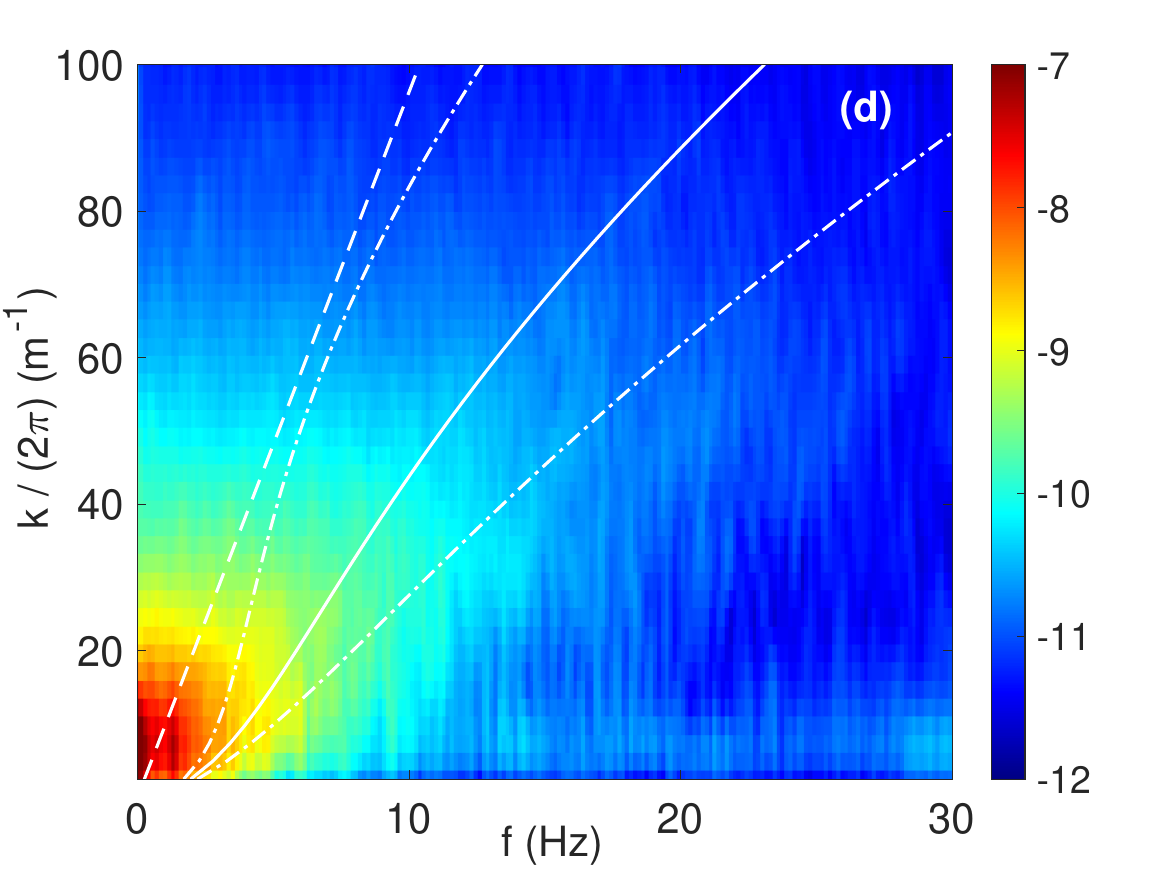}%Fig5bN.pdf}
\caption{Spatiotemporal power spectra $S_\eta(k, f)$ of free-surface deformations for different subsurface turbulence: weak (a) $\sigma_{\mathcal{U}} = 0.65$~cm/s and (b) $\sigma_{\mathcal{U}} = 1.24$~cm/s, (c) moderate turbulence $\sigma_{\mathcal{U}} = 5.1$~cm/s, and (d) strong turbulence $\sigma_{\mathcal{U}} = 10.4$~cm/s. Logarithmic-scale colorbars in m$^3$~s. White solid curves: dispersion relation of Eq.~(\ref{eq:dispersion_doppler}) with $\sigma_{\mathcal{U}}=0$, i.e., without large-scale random flow advection. White dashed lines: $\omega = k_x \sigma_{\mathcal{U}}$ (pure advected structures). White dash-dotted curves: Doppler-shifted dispersion relation of Eq.~(\ref{eq:dispersion_doppler}) (advected gravity-capillary waves) with corresponding values of $\sigma_{\mathcal{U}}$. }
\label{fig:spatiotemporal_weak}
\end{figure}

\subsection{Spatiotemporal spectrum of surface deformations}
To elucidate the physical nature of surface deformations, we compute the spatiotemporal power spectrum $S_\eta(k, f)$ of the free-surface deformations. Such a spectrum reveals whether the surface exhibits propagating waves (concentrated along dispersion relation curves) or non-propagating advected structures (concentrated near $f\simeq 0$).  Due to the isotropy of the deformations, the spatiotemporal power spectrum of the free-surface deformations, $S_\eta(k,\omega)$, is obtained as $S_\eta(k,\omega)=\int_kS_\eta(k_x,k_y,\omega)dk_xdk_y$ with $k\equiv\sqrt{k_x^2+k_y^2}$ and $S_\eta(k_x, k_y, \omega)\equiv |\int \hat{\eta}(k_x,k_y,t)e^{i\omega t}dt|^2/(\sqrt{2\pi}\mathcal{ST})$. $\hat{\eta}(k_x,k_y,t)\equiv\int \eta(x,y)e^{-i(k_xx+k_yy)}dxdy/(2\pi)$ is the 2D spatial Fourier transform of the surface deformation amplitude $\eta(x,y)$ of each image recorded at time $t$. %When integrating the spectrum over angles, one has   

%\subsubsection{Gentle turbulence}
Figure~\ref{fig:spatiotemporal_weak} shows the spatiotemporal spectra, $S_\eta(k,\omega)$, for different turbulence levels $\sigma_{\mathcal{U}}$. For gentle turbulence (Fig.~\ref{fig:spatiotemporal_weak}a-b), we observe an energy concentration near $f\simeq 0$ in the spectra, whereas for moderate (Fig.~\ref{fig:spatiotemporal_weak}c) and strong (Fig.~\ref{fig:spatiotemporal_weak}d) turbulence, energy broadening occurs in both frequency and wavenumber. To interpret these spectra, we must account for the Doppler shift due to random large-scale advection beneath the surface (although almost no bulk mean flow is present~\cite{Jamin2025}). For gravity-capillary waves propagating on a medium in the presence of random large-scale advection of strength $\sigma_{\mathcal{U}}$, the dispersion relation reads
\begin{equation}
\omega = \sqrt{gk + \frac{\gamma k^3}{\rho}} \pm k \sigma_{\mathcal{U}}
\label{eq:dispersion_doppler}
\end{equation}
where $g$ is gravitational acceleration, $\gamma$ is surface tension, and $\rho$ is water density. The white dash-dotted curves in Fig.~\ref{fig:spatiotemporal_weak} correspond to Eq.~(\ref{eq:dispersion_doppler}) and the white solid ones to Eq.~(\ref{eq:dispersion_doppler}) with $\sigma_{\mathcal{U}}=0$, i.e., without random large-scale advection. For gentle turbulence (Fig.~\ref{fig:spatiotemporal_weak}a-b), no gravity-capillary wave is generated and a significant portion of energy concentrates along the line $\omega = k \sigma_{\mathcal{U}}$ (white dotted lines in Fig.~\ref{fig:spatiotemporal_weak}), corresponding to random large-scale advection. This energy dominates over that of propagating waves, contrasting with the results of Savelsberg and van de Water (turbulent current with a strong mean flow in an open channel)~\cite{Savelsberg2009} but consistent with numerical simulations by Guo and Shen~\cite{Guo2010} for similar configurations (no mean flow and finite domain). 

It should be noted that if the spectrum is computed along a single horizontal direction (i.e., not integrated over all directions), the gravity-capillary surface-wave dispersion relation can be shifted by a mean flow (not shown). This occurs for weak forcing, where the fully developed turbulence regime is not achieved, and for measurement durations shorter than those in~\cite{Jamin2025}. However, this mean-flow effect is not dominant compared to that of surface fluctuations, as the shifted dispersion-relation branch is no longer visible when the spectrum is integrated over all directions.
%to low-frequency structures advected by random large-scale flow. These structures, directly related to subsurface turbulence, could be upwelling events or surface scars as described by Brocchini and Peregrine~\cite{Brocchini2001}. Importantly, the energy associated with these low-frequency advected structures

For strong enough turbulence (Fig.~\ref{fig:spatiotemporal_weak}c-d), part of the spectral energy distribution is located around the dispersion relation curve, indicating the presence of gravity-capillary waves generated by the turbulent flow. However, the energy spreads more broadly due to Doppler effects: The velocity fluctuations of turbulent eddies advecting waves create a frequency broadening for each wavenumber. This spectral broadening is thus different from the classical nonlinear broadening of weak wave turbulence occurring only near the dispersion relation~\cite{FalconARFM2022,RicardEPL2021}. Here, the Doppler broadening from turbulent velocity fluctuations increasingly spreads spectral energy both spatially and temporally. This makes it difficult to distinguish between energy from propagating waves (which would concentrate along dispersion curves without advection) and low-frequency advected structures (which would concentrate near $f\simeq 0$ without advection). 

The substantial energy spreading observed in frequency suggests large subsurface velocities near the interface and significant wave presence. However, as we show next through spatial and temporal spectral analyses, the evidence points to the continued dominance of advected low-frequency structures across all turbulence levels.

\begin{figure}[t!]
\centering
\includegraphics[width=0.45\textwidth]{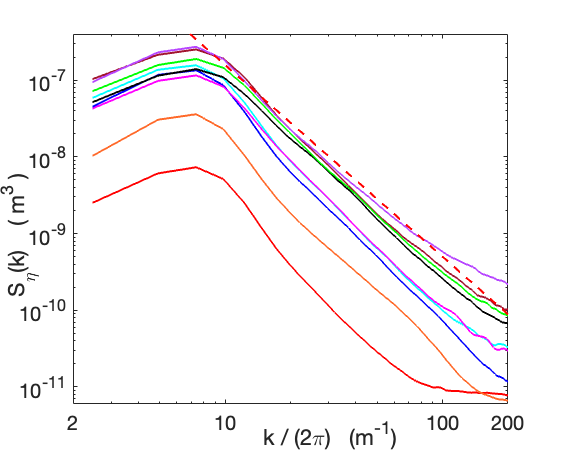} %
\caption{Spatial power spectra $S_\eta(k)$ for different turbulence intensities $\sigma_{\mathcal{U}} = 0.65$, 1.2, 2.4, 3.7, 5.1, 6.9, 8.3, and 10.4~cm/s (from bottom to top). The red dashed line shows the power law $k^{-2.5}$ in the intermediate range. All spectra exhibit similar shapes with consistent power-law exponents. The integral scale of the subsurface turbulence is $1/L\approx 12.5$~m$^{-1}$.}
\label{fig:spatial}
\end{figure}

\subsection{Spatial spectrum of surface deformations}\label{spacespec}
The spatial power spectrum, $S_\eta(k)\equiv\int S_\eta(k,\omega)d\omega$, is obtained by integrating the spatiotemporal spectrum over all frequencies. Figure~\ref{fig:spatial} shows such spectra for all turbulence intensities $\sigma_{\mathcal{U}}$ investigated, and reveals the spreading of surface deformation energy across spatial scales. All spectra exhibit remarkably similar shapes with a maximum at $k_{\rm max}/(2\pi) \approx 7.4$~m$^{-1}$ (i.e., $k_{\rm max} \approx 46$~m$^{-1}$ or wavelength $\lambda_{\rm max} \approx 13.5$~cm), slightly larger than the integral length scale $L \approx 8$~cm, as found for the velocity spectrum below the surface in the homogeneous region~\cite{Jamin2025}. For intermediate scales ($k/(2\pi)\in [10, 100]$~m$^{-1}$ or $\lambda \in [1,10]$~cm, corresponding to the gravity-capillary wave regime), the spectra follow a power law:
\begin{equation}
S_\eta(k) \propto k^\alpha
\end{equation}
over one decade in $k$ with exponent $\alpha = -2.5\pm 0.1$ consistent across all turbulence levels. This power-law exponent differs markedly from that ($\alpha \approx -6$) found when surface dynamics is dominated by waves in the presence of subsurface turbulence in a channel with a mean flow~\cite{Savelsberg2009}. Instead, our value is consistent with studies for quasi-two-dimensional turbulent flows where no waves were observed ($\alpha$ between $-2$ and $-3.5$)~\cite{Gutierrez2013}. The spatial exponent $\alpha \approx -2.5$ is close to theoretical predictions for several scenarios. For gravity waves, weak turbulence theory predicts $\alpha = -5/2=-2.5$~\cite{Zakharov1967,Zakharov2025,FalconARFM2022}, and at stronger forcing, a saturated spectrum (flux-independent solution) predicted by Phillips with $\alpha=-3$~\cite{Phillips1958}, both compatible with field observations in ocean~\cite{MelvilleJAOT2016}; whereas for capillary waves, weak turbulence predicts an exponent $-15/4=-3.75$~\cite{Zakharov1967b,Zakharov2025,FalconARFM2022}. However, our spatiotemporal spectra at gentle turbulence clearly showed that low-frequency advected structures dominate over propagating waves. A plausible alternative interpretation is that surface deformations passively respond to subsurface turbulent pressure fluctuations. From hydrostatic balance, $\eta \approx p/(\rho g)$ where $p$ represents pressure fluctuations. The spatial structure of turbulence-induced surface deformations should therefore mirror the structure of the subsurface pressure field. Model and experiments show that the spatial spectrum of a turbulent pressure field $p$ scales as $S_p(k) \propto k^{-7/3}$ at sufficiently high Reynolds numbers~\cite{Monin1975,Tsuji2003,Terashima2012}, close to our observed $k^{-2.5}$ scaling. The consistent power-law exponent and spectral shape across all turbulence intensities suggest that the same deformation mechanism operates throughout our parameter range, supporting the passive response hypothesis.

\begin{figure}[t!]
\centering
\includegraphics[width=0.45\textwidth]{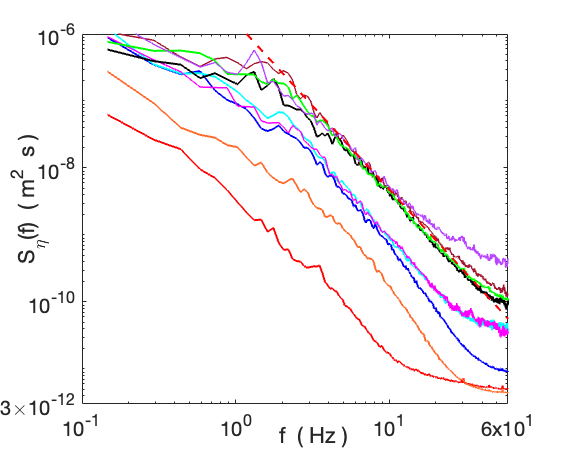}
\caption{Frequency power spectra $S_\eta(f)$ for different turbulence intensities $\sigma_{\mathcal{U}} = 0.65$, 1.2, 2.4, 3.7, 5.1, 6.9, 8.3, and 10.4~cm/s (from bottom to top). The red dashed line shows the power law $f^{-2.5}$ in the inertial range. The similarity with spatial spectral exponents suggests advection-dominated dynamics.}
\label{fig:temporal}
\end{figure}

\subsection{Frequency spectrum of surface deformations}
Frequency power spectra, $S_\eta(\omega)\equiv \int S_\eta(k,\omega)dk$, obtained by integrating spatiotemporal spectra over all wavenumbers, characterize the frequency content of surface deformations. Figure~\ref{fig:temporal} shows these spectra for all turbulence levels. For intermediate turbulence intensities (excluding the two weakest), the inertial range exhibits a power law:
\begin{equation}
S_\eta(f) \propto f^\beta
\end{equation}
with $\beta = -2.5\pm 0.1$ across these cases. For propagating waves, spatial and temporal spectra are related through the dispersion relation $\omega\sim k^{\zeta}$. Specifically, if $S_\eta(k) \propto k^\alpha$ and $S_\eta(f) \propto f^\beta$, dimensional analysis then leads to $\beta = (\alpha + 1)/\zeta-1$~\cite{Zakharov2025}. Thus, for gravity waves ($\omega = \sqrt{gk}$, i.e., $\zeta=1/2$) and for $\alpha = -2.5$, one should have $\beta = -4$, and weak gravity-wave turbulence theory indeed predicts $\beta = -4$~\cite{Zakharov1967,Zakharov2025} as well as Phillips' theory for wind-generated resonant waves~\cite{Phillips1985}, whereas Phillips' saturated spectrum predicts $\beta = -5$~\cite{Phillips1958}. For capillary waves ($\omega = \sqrt{\gamma k^3/\rho }$, i.e., $\zeta=3/2$), $\alpha = -2.5$ should lead to $\beta \approx -2$ whereas weak capillary-wave turbulence predicts $\beta=-17/6 \approx -2.8$ and $\alpha=-15/4=-3.75$~\cite{Zakharov1967b,Zakharov2025}. However, here, we observe $\beta \approx -2.5 \approx \alpha$, indicating that temporal and spatial power-law exponents are nearly equal rather than related by the dispersion relationship. This similarity can be explained if low-frequency structures are advected by the flow. For such advected structures with $\omega \approx kv$ (i.e., $\zeta=1$) where $v$ is a characteristic velocity of the subsurface flow, we have $d\omega/dk \approx v$, leading to $\beta \approx \alpha$. We verified that subsurface turbulent structures satisfy Tennekes' frozen turbulence hypothesis in our flow, consistent with advection dominating over intrinsic dynamics for turbulent eddies~\cite{Jamin2025}. The concordance between experimental spatial and temporal power-law exponents ($\beta \approx \alpha$) therefore supports the hypothesis that low-frequency advected structures dominate surface deformations, rather than propagating waves, across most turbulence levels studied.  

\section{Model of passive response to turbulent pressure fluctuations}
To further investigate the passive response mechanism and describe our experimental observations, we develop a model for the linear response of a liquid surface submitted to an arbitrary normal $N$ and tangential $\mathbf{T}$ stress fields in a statistically steady and homogenous regime, following the approach of Perrard et al.~\cite{Perrard2019}.  After linearization of the Navier-Stokes equations with such free-surface boundary conditions, the spatiotemporal Fourier transform of the surface elevation reads~\cite{Perrard2019}
\begin{equation}
\hat{\eta}(\mathbf{k},\omega) = \frac{k \hat{N} + i  \mathbf{k} \cdot \hat{\mathbf{T}}}{\rho (\omega^2 - g k - \gamma k^3/\rho + i 2\omega \delta)}
\label{eq:transfer_function}
\end{equation} 
where the denominator represents the transfer function of the free surface, and $\hat{N}$ and $\hat{\mathbf{T}}$ are the Fourier components of the applied stresses. The real part vanishes when the forcing satisfies the surface wave dispersion relation $\omega(k)$, leading to resonance that is saturated by the dissipative term [imaginary part involving a wave damping rate $\delta(\omega)$]. %for uncontaminated water $\delta_0 = 2\nu k^2$.

We evaluate the free-surface response to turbulent pressure fluctuations in the liquid phase below the free surface, assuming negligible stresses in the gas phase. For homogeneous isotropic turbulence, the 1D pressure spatial power spectrum is predicted to follow~\cite{Monin1975}
\begin{equation}
S_P^{\rm 1D}(k)=C \rho^2 \epsilon^{4/3} k^{-7/3}
\label{eq:pressure_spectrum}
\end{equation}
where $\epsilon$ is the turbulent energy dissipation rate, and $C=3.59$ is a constant. This power law is verified experimentally~\cite{Tsuji2003,Terashima2012}, and the constant is close to $8\pm0.5$ numerically~\cite{GotohPRL2001}. Although turbulence becomes anisotropic near the free surface~\cite{Jamin2025}, pressure $P$ is a nonlocal field resulting from integration of velocity gradient source terms with $1/r$ decay (Poisson equation inversion). %\rev{The pressure field near a boundary indeed results from homogeneous isotropic turbulence as observed experimentally~\cite{FauveJphys1993} (for us, typically deeper than $-6$~cm).} 
As normal stress forcing $N$ strongly dominates tangential stress $\mathbf{T}$ in turbulence~\cite{Perrard2019}, we assume $N = P$ and $T = 0$. For the pressure spectrum $S_P(k,\omega)$, we use the random sweeping model assuming that the small-scale fluctuations in a turbulent flow are randomly swept by the large-scale fluctuations~\cite{Wilczek2012}. The spatiotemporal spectrum consists of a spatial spectrum multiplied by a Gaussian frequency distribution, reading (when averaged over horizontal directions)~\cite{Wilczek2012} 
\begin{equation}
S_P(k,\omega) = \frac{S_P^{\rm 1D}(k)}{\sqrt{2\pi k^2 V^2}} \exp\left[-\frac{(\omega - kU)^2}{2 k^2 V^2}\right]
\label{eq:sweeping_model}
\end{equation}
with $U$ the mean flow velocity, $V = \sigma_{\mathcal{U}}/\sqrt{3}$ the sweeping velocity, and $\sigma_{\mathcal{U}}$ the rms amplitude of turbulent velocity fluctuations. This model has interesting properties. The mean velocity leads to a Doppler shift in frequencies, whereas the sweeping velocity broadens the spectrum in the frequency domain. From integration of Eq.~\eqref{eq:sweeping_model}, the frequency spectrum exhibits the same spectral power-law exponent as the spatial spectrum, independent of the mean and sweeping velocities. Taking the square modulus of Eq.~\eqref{eq:transfer_function}, the power spectrum of surface deformations $S_\eta(k,\omega)$, as the response to $S_P(k,\omega)$ of Eq.~\eqref{eq:sweeping_model}, is given by
\begin{equation}
S_\eta(k,\omega) = \frac{k^2 S_P(k,\omega)}{\rho^2 |\omega^2 - g k - \gamma k^3/\rho + i 2\omega \delta|^2}
\label{eq:surface_response}
\end{equation}

A crucial aspect is that turbulence strongly attenuates the waves it might create. We use the damping rate due to turbulent advection, $\delta_u = \sigma_{\mathcal{U}} k (L k)^{-1/3}$, with $L$ the integral scale~\cite{Gutierrez2016}. This term generally dominates viscous damping $\delta_{\nu_0} = 2\nu k^2$ (for uncontaminated surface case), and even for fully contaminated surfaces where $\delta_\nu = k\sqrt{\nu \omega}/(2\sqrt{2})$~\cite{Lamb1932,DeikePRE2014}. We take $\delta = \delta_u + \delta_\nu$, representing a typical inverse lifetime of surface fluctuations.

\begin{figure}[ht!]
\includegraphics[width=0.45\textwidth]{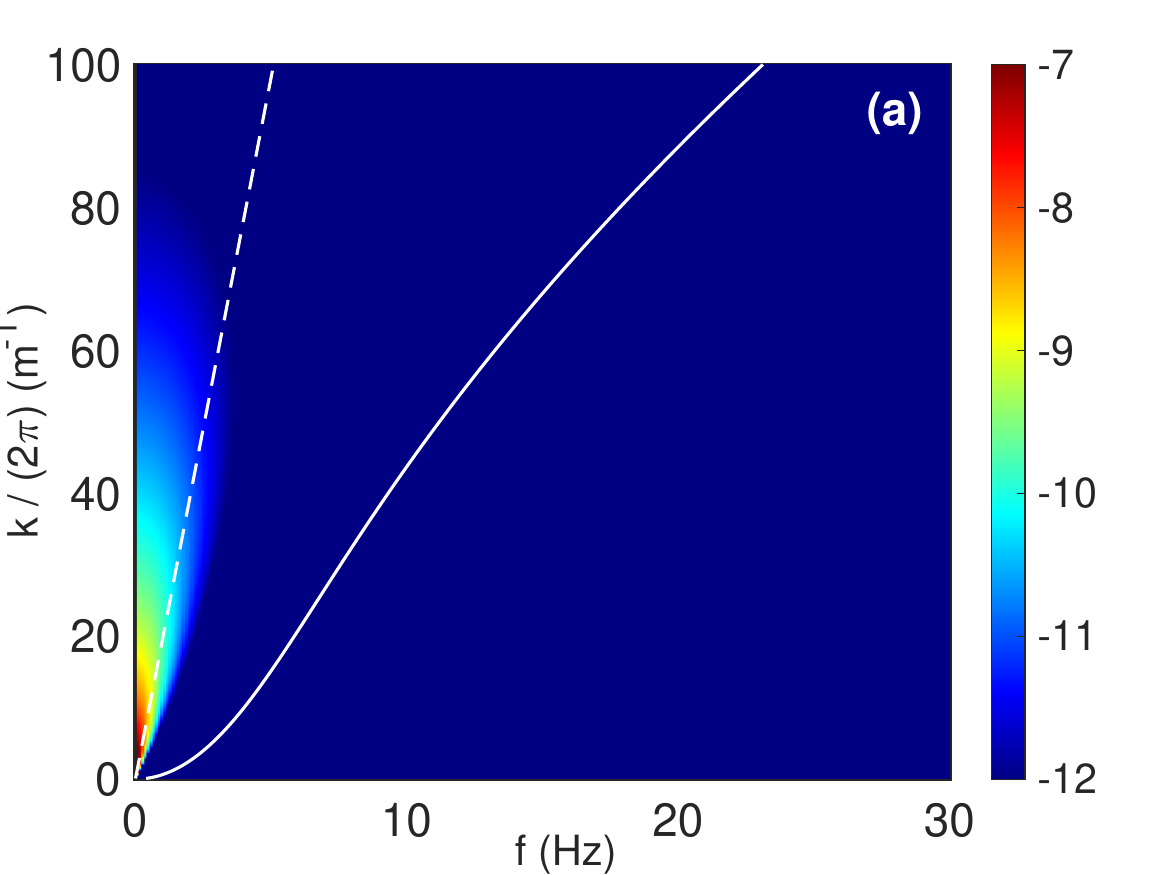}%Fig8a-eps-converted-to.pdf}%PSD-eta5cms-eps-converted-to.pdf}
\includegraphics[width=0.45\textwidth]{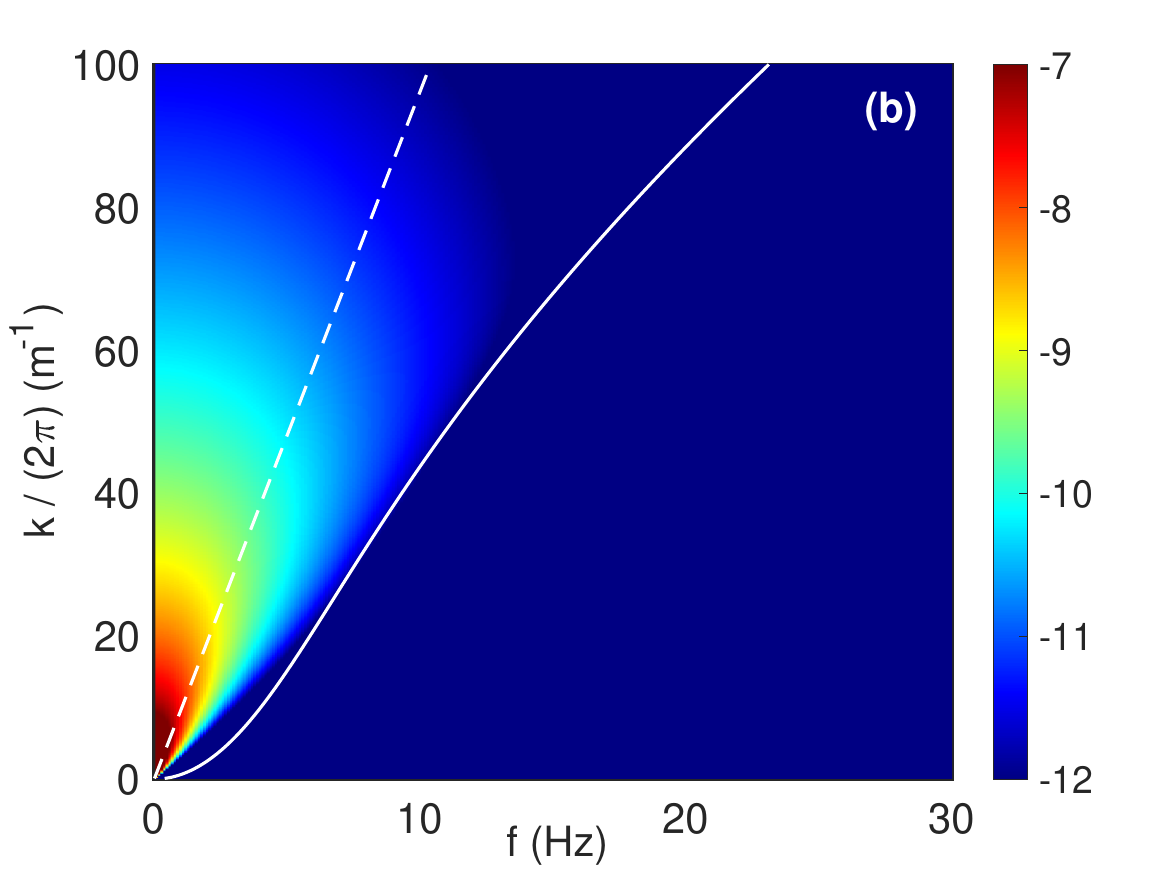}\\%Fig8b-eps-converted-to.pdf}\\%PSD-eta10cms-eps-converted-to.pdf}
\includegraphics[width=0.45\textwidth]{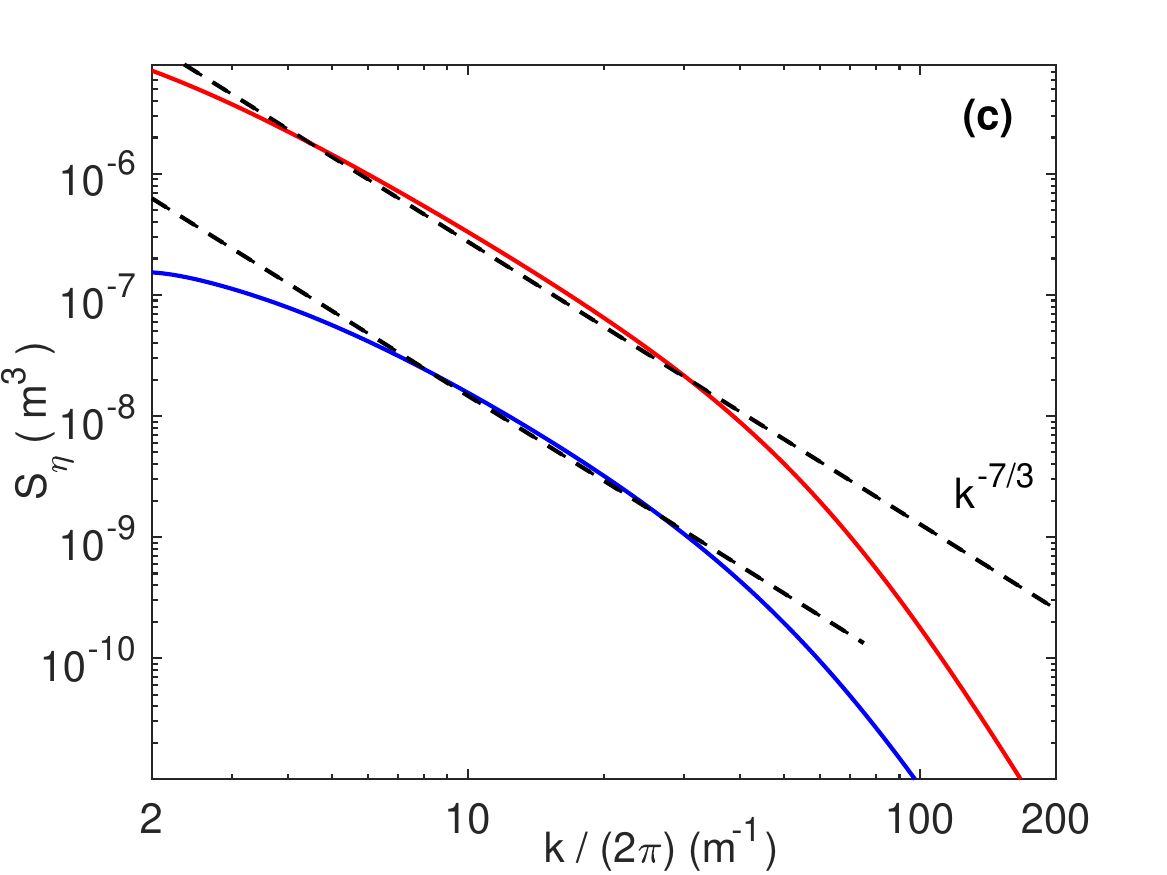}%SpspaN-eps-converted-to.pdf}
\includegraphics[width=0.45\textwidth]{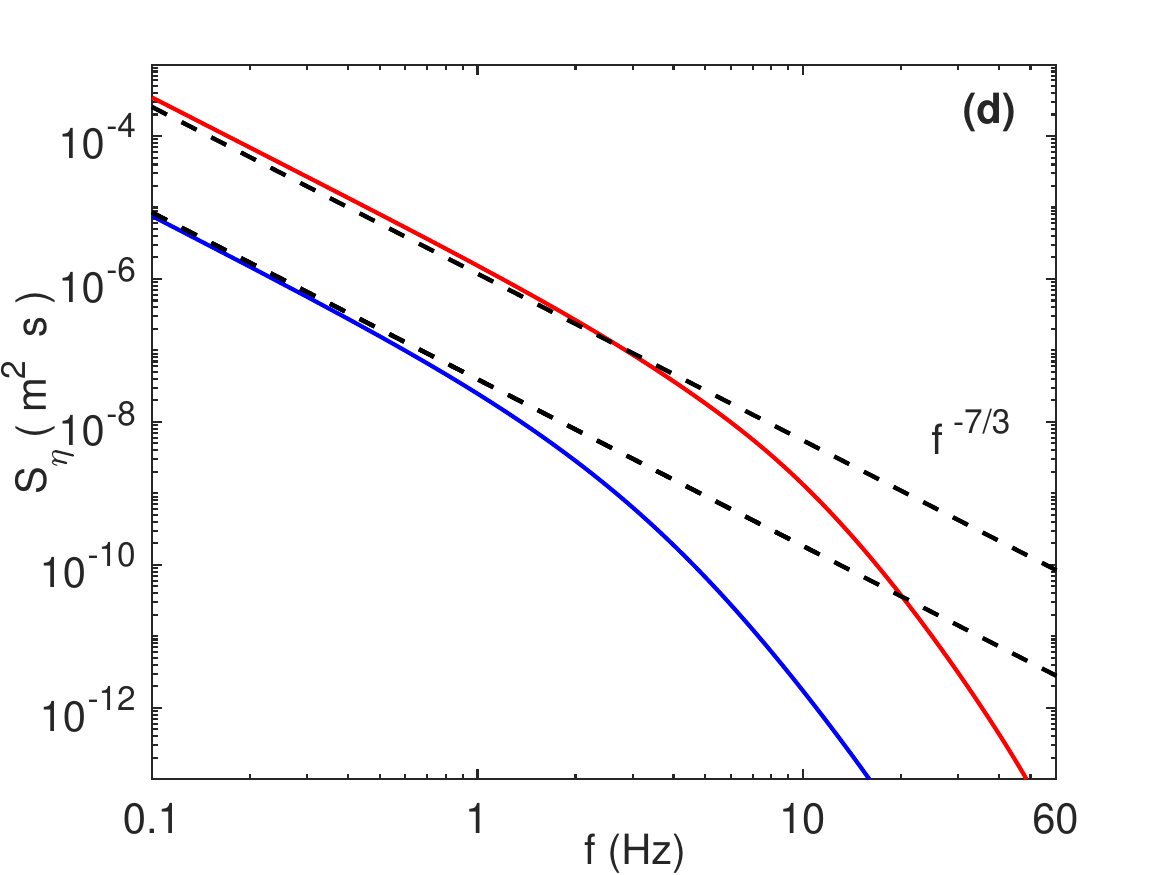}\\%SptN-eps-converted-to.pdf}
\caption{Model predictions: Spatiotemporal spectra of surface deformations $S_\eta(k,\omega)$ for (a) weak subsurface turbulence ($\sigma_{\mathcal{U}} =5.1$~cm/s) and (b) strong turbulence ($\sigma_{\mathcal{U}} =10.4$~cm/s). Logarithmic-scale colorbars in m$^3$~s. White dotted lines: $\omega =\sigma_{\mathcal{U}}k$. White curves show the gravity-capillary dispersion relation of Eq.~\eqref{eq:dispersion_doppler} with $\sigma_{\mathcal{U}} =0$.  Model predictions: Spatial (c) and temporal (d) power spectra of surface deformation for (blue) weak subsurface turbulence ($\sigma_{\mathcal{U}}=5.1$~cm/s) and (red) strong turbulence ($\sigma_{\mathcal{U}}=10.4$~cm/s). The dashed lines are $k^{-7/3}$ (c) and $f^{-7/3}$ (d) power laws.} %Experiments: Spatiotemporal spectra of surface deformations $S_\eta(\omega,k)$ for (c) weak turbulence ($\sigma_{\mathcal{U}}=5$~cm/s) and (d) strong turbulence ($\sigma_{\mathcal{U}}=10$~cm/s).  
\label{fig:model}
\end{figure}

Using our experimental parameters ($L = 8$~cm, $\epsilon = A\sigma_{\mathcal{U}}^3/L$, A=0.42, no mean flow $U=0$)~\cite{Jamin2025} and Eq.~\eqref{eq:pressure_spectrum} with $C=3.59$, we compute predictions of Eqs.~\eqref{eq:sweeping_model} and \eqref{eq:surface_response} for two turbulent velocity fluctuations $\sigma_{\mathcal{U}}$. First, we check that the pressure spectrum is consistent with previous direct numerical simulations~\cite{ClarkDiLeoni2015}. Figure~\ref{fig:model}a then shows the surface deformation spectrum for weak turbulence ($\sigma_{\mathcal{U}} = 5$~cm/s). The energy is mainly concentrated in the region $\omega \leq \sigma_{\mathcal{U}} k$, and energy is absent on the dispersion relation (white line). These features closely resemble the pressure fluctuation spectrum, indicating a passive response (not shown). For stronger turbulence ($\sigma_{\mathcal{U}} = 10$~cm/s), the surface spectrum extends over a broader ($\omega$,$k$) region (see Fig.~\ref{fig:model}b), qualitatively resembling our experimental spatiotemporal spectra in Fig.~\ref{fig:spatiotemporal_weak}. Surface waves could be excited, but the damping, $\delta_u$, due to turbulent advection remains too strong. If we neglect this term and retain only viscous dissipation, $\delta_\nu$, (contaminated surface case), much stronger amplitudes appear along the wave dispersion relation (not shown).

By integrating the spatiotemporal spectrum $S_\eta(k,\omega)$ of Eq.~\eqref{eq:surface_response} in $\omega$ or in $k$, the predicted spectra in Fig.~\ref{fig:model}c-d show power laws $S_\eta(k) \propto k^{-7/3}$ and $S_\eta(f) \propto f^{-7/3}$ in the gravity-capillary wave range in good agreement with experimental spectra of Figs.~\ref{fig:spatial} and \ref{fig:temporal}, respectively.  %where the transfer function is essentially flat.
Finally, the predicted rms surface fluctuations are inferred from $\sigma_\eta \equiv \sqrt{\int S(k)dk}$ leading to $\sigma_\eta \approx 1.2$~mm (resp. 0.2~mm) for $\sigma_{\mathcal{U}} = 10.4$~cm/s (resp. 5.1~cm/s) in rough agreement (given the simplifying assumptions) with the measured values $2.1$~mm (resp. $1.1$~mm) in Fig.~\ref{fig:amplitude}a. 
%$\sigma_\eta \equiv \sqrt{\int S(k)dk} \approx 0.29$~mm (or 1.16~mm) for $\sigma_{\mathcal{U}} = 5$~cm/s (or 10~cm/s), roughly less than half the measured value but within the correct order of magnitude given the simplifying assumptions.
%The fluctuations scale as $\sigma_{\mathcal{U}}^2$ in this regime.

Our model thus successfully captures several key experimental observations. First, the spatiotemporal spectra show energy concentration at low frequencies in the region $\omega \leq \sigma_{\mathcal{U}} k$ rather than along the dispersion relation, consistent with advection-dominated dynamics. Second, spatial and temporal power-law exponents are similar, matching our experimental observation that $\alpha \approx \beta$ rather than the wave relation $\beta = (\alpha + 1)/\zeta-1$. Third, the predicted exponent $-7/3 \approx -2.33$ is close to our experimental value $-2.5\pm0.1$, confirming passive response to turbulent pressure fluctuations. Moreover, the model confirms that turbulent damping $\delta_u$ prevents energy concentration along the dispersion relation, explaining why wave signatures remain weak in our experiments. The passive response mechanism thus creates randomly advected structures that carry the signature of subsurface pressure fluctuations. It is worth noting, however, that the model predicts $\sigma_{\eta}\propto \sigma_{\mathcal{U}}^2$, in contrast with the linear dependence observed experimentally. Indeed, contributions to the free-surface fluctuations in the experiment could be due to other phenomena, such as large-scale coherent flow structures of upwelling or downwelling~\cite{Guo2010}, thus not taken into account in our model.
Minor quantitative discrepancies are also observed as small differences in the spectral slopes, and energy is more spread out experimentally near the dispersion relation. They likely arise from simplified assumptions in the model (homogeneous isotropic pressure spectrum near the anisotropic surface, and empirical turbulent damping formula) and from experimental precision to distinguish between $-7/3$ and $-2.5\pm0.1$ exponents on a single decade scale. However, the model captures the main physics and validates our interpretation of surface deformations as resulting partially from passive responses to turbulent pressure~fields.

\section{Discussion}
Our experimental investigation reveals that passive response to subsurface pressure fluctuations contributes significantly to surface deformations. The experimental spectra in Fig.~\ref{fig:spatiotemporal_weak} display low amplitude close to the surface wave dispersion relation, even when accounting for broadening due to turbulent fluctuations, revealing a small contribution of coherent surface waves. The passive response mechanism creates advected structures carrying the signature of turbulent pressure fields, with a spectral exponent ($-2.5$) for the surface deformation spectrum matching the model prediction ($-7/3$). The coexistence of passive response and weak propagating waves distinguishes our results from two limiting cases in the literature. Savelsberg and van de Water~\cite{Savelsberg2009} observed wave-dominated surfaces in grid-stirred turbulence with a strong mean flow and a lower turbulence intensity.  A clear wave signature, with a much steeper $k$-spectral slope ($-6$), is observed, without highlighting the wave creation mechanism. Conversely, Guo and Shen's simulations~\cite{Guo2010} of jet-forced turbulence without mean flow showed structure-dominated surfaces similar to ours, supporting the importance of configuration geometry in determining the dominant mechanism. The turbulent damping rate $\delta_u \sim \sigma_{\mathcal{U}} k (Lk)^{-1/3}$ identified by Gutiérrez and Aumaître~\cite{Gutierrez2016} should play an important role in our system. Our model demonstrates that including $\delta_u$ is essential to reproduce the weak wave activity arising from the passive response to turbulent pressure fluctuations. Without turbulent damping, the model predicts much stronger wave signatures along the dispersion relation. However, our model does not describe the linear scaling $\sigma_\eta \propto \sigma_{\mathcal{U}}$ found experimentally across our range of turbulence intensities, possibly due to the large-scale coherent flow structures (such as upwelling and downwelling) characteristics of free-surface turbulence~\cite{Guo2010}. This observed linear scaling contrasts with free-surface breaking regimes where nonlinear effects become important at higher Froude numbers~\cite{Calado2025}. Our moderate Froude numbers maintain the free surface in a quasilinear response regime, validating the linear transfer function approach. Recent work on jet-forced turbulence focused on subsurface turbulence structure rather than surface deformations, but their observation of strong upwelling/downwelling events is consistent with our interpretation~\cite{Ruth2024,Qi2025}. A linear response to a transient vertical current $U_z$ can be described by the linear kinematics boundary condition $\partial \eta / \partial t \approx U_z$. Assuming $U_z$ scales as $\sigma_{\mathcal{U}}$, this condition may explain the linear scaling between $\sigma_\eta$ and $\sigma_{\mathcal{U}}$. We note indeed Fig.~\ref{fig:amplitude}a that the coefficient $K$ of Eq.~\eqref{EtaUscaling} of dimension of a time, is of order $\sqrt{L/g}$. $\sigma_{\mathcal{U}}$ corresponds indeed to the turbulent fluctuations at the integral scale $L$.  This linear response to transient turbulent currents may represent the primary mechanism populating low-frequency, large-scale components of the spectra, whereas the response to pressure fluctuations is likely responsible for the observed power-law behavior. A more elaborate model combining kinetic and dynamic boundary conditions would enable a more complete description of turbulent free-surface dynamics. The coupling between large-scale coherent transient currents and turbulent fluctuations drives the passive response of the free surface. Joint measurements of surface deformations and subsurface velocity fields would enable a direct quantification of this coupling. 

\section{Conclusion}
We have reported the characterization of spatiotemporal deformations of a free surface forced by 3D homogeneous and isotropic turbulence with zero mean flow. Over a wide range of turbulence intensities, we found that the surface deformation amplitude scales linearly with subsurface velocity fluctuations ($\sigma_\eta \propto \sigma_{\mathcal{U}}$), up to a significant steepness ($\approx 0.1$), without exhibiting threshold behavior across the entire range. Surface deformations are strongly advected by subsurface flows. Their spatiotemporal spectra reveal the coexistence of two phenomena. First, large-scale coherent structures (upwelling) provide low-frequency surface deformation. Second, the passive response to subsurface turbulent pressure fluctuations produces at the free-surface the footprint of the pressure spectrum filtered by the linear transfer function of the free-surface. Spatial and temporal spectra of surface deformations exhibit similar wavenumber- and frequency-power-law exponents ($-2.5$), consistent with advection-dominated dynamics rather than free wave propagation. These spectral characteristics remain remarkably consistent across our range of turbulence levels. We also developed a linear response model based on the transfer function of the free surface to turbulent pressure fluctuations below, involving turbulent wave damping that prevents wave amplification. The model successfully predicts the main features of the free-surface dynamics: (i) spatiotemporal spectrum shape with energy concentrated at low-frequency in the region $\omega \leq \sigma_{\mathcal{U}} k$, (ii) power-law exponents $k^{-7/3}$ and $\omega^{-7/3}$ matching turbulent pressure spectra, and (iii) dominance of passive response over free wave generation. The $-7/3$ spectral-exponent prediction aligns well with experimental observations ($-2.5$), confirming that surface deformations primarily reflect the structure of the subsurface turbulent pressure field. In configurations lacking conditions for Phillips resonance, i.e., specifically when turbulent velocities remain below wave phase velocities ($\sigma_{\mathcal{U}} < c^{\rm min}_{w}$) and interfacial shear is weak (e.g., no mean flow), surface deformations are dominated by passive response to the turbulent pressure fluctuations beneath the free surface. The turbulent damping rate prevents significant wave generation even as forcing increases. This regime, reported here, thus differs from wave-dominated surfaces observed in grid-stirred turbulence with a strong mean flow (while remaining far from the resonance conditions for wave amplification)~\cite{Savelsberg2009}, highlighting the critical importance of flow configuration. These results complement recent advances in free-surface turbulence~\cite{Ruth2024,Bullee2024,Jamin2025,Calado2025,Qi2025,AarnesJFM2025,BabikerArXiv2025} and provide quantitative validation of the passive response mechanism in the nonresonant regime. This is particularly relevant to many environmental and industrial flows where wave generation conditions are not met, including regions of the ocean away from strong wind forcing, industrial mixing tanks, and flows in confined geometries. Future work could explore different forcing configurations satisfying resonance conditions to map the transition between structure- and wave-dominated regimes. New forcing method of turbulence at small scale, using magnetic stirrers randomly driven in space and time, could also be tested in this context~\cite{FalconPRF2017,Cazaubiel2021,Gorce2022}. Other work could also examine the wave generation mechanism (possibly by upwellings or transient humps), and finally extend measurements to higher turbulence approaching boiling and bubble formation regimes, requiring adapted optical techniques. 

\section*{Acknowledgments}
We thank T. Jamin for insightful discussions and early contributions to this work. We thank A. Lantheaume, Y. Le Goas, and A. Di Palma for technical support. This work was supported by the French National Research Agency (ANR TURBULON project No. ANR-12-BS04-0005 and ANR DYSTURB project No. ANR-17-CE30-0004), the DGA (Direction Générale de l’Armement, France, No. 2015600018), and the CNRS. E.~F. thanks the support of the Simons Foundation Project No. MPS-WT-00651463 (U.S.), ANR Lascaturb Project No. ANR-23-CE30-0043-02, and ANR Provebact Project No. ANR-24-CE09-1394-02.

\end{document}